\documentclass[12pt]{article}
\usepackage{epsfig,amssymb,array}
\usepackage{amsmath}
\usepackage{graphics,psfrag,rotating,cite}

\def\kslash{{k\mkern-7mu/}}

\def\partials{{\partial \mkern-9mu/}}
\def\lsim{\raisebox{-.4ex}{$\stackrel{<}{\scriptstyle \sim}$\,}}
\def\gsim{\raisebox{-.4ex}{$\stackrel{>}{\scriptstyle \sim}$\,}}

\makeatletter
\renewcommand{\section}{\setcounter{equation}{0}\@startsection
    {section}%
    {1}%
    {0pt}%
    {-1\baselineskip}%
    {0.4\baselineskip}%
    {\large\bfseries}}%
\renewcommand{\subsection}{\@startsection
    {subsection}%
    {2}%
    {0pt}%
    {-0.75\baselineskip}%
    {0.2\baselineskip}%
    {\bfseries}}%
\renewcommand{\subsubsection}{\@startsection
    {subsubsection}%
    {3}%
    {0pt}%
    {-0.5\baselineskip}%
    {0.1\baselineskip}%
    {\sc}}%
\makeatother

\begin{document}
\input epsf \renewcommand{\topfraction}{0.8}
\pagestyle{empty} \vspace*{-5mm}
\begin{center}
\Large{\bf Towards the effective potential of the Littlest Higgs model } \\
\vspace*{2cm} \large{ Antonio Dobado, Lourdes Tabares} \\
\vspace{0.2cm} \normalsize
Departamento de  F\'{\i}sica Te\'orica I,\\
Universidad Complutense de
Madrid, E-28040 Madrid, Spain
\vspace{0.2cm}\\
and\vspace{0.2cm}\\
\large{Siannah Pe\~naranda}
\\ \vspace{0.2cm} \normalsize
Departamento de F{\'{\i}}sica Te{\'o}rica, Universidad
de Zaragoza, E-50009, Zaragoza, Spain\\
\vspace*{2cm}{\bf ABSTRACT}
\end{center}
We compute the relevant parameters of the combined Higgs and
$\phi$ scalar effective potential in the Littlest Higgs (LH)
model. These parameters are obtained as the sum of two kind of
contributions. The first one is the one-loop radiative corrections
coming from fermions and gauge bosons. The second one is obtained
at the tree level from the higher order effective operators needed
for the ultraviolet completion of the model. Finally we analyze
the restrictions that the requirement of reproducing the standard
electroweak symmetry breaking of the SM set on the LH model
parameters.
 \noindent
\newpage
\setcounter{page}{1} \pagestyle{plain} \textheight 20 true cm

\section{Introduction}

The discovery of a Higgs boson and the elucidation of the
mechanism responsible for the electroweak symmetry breaking are
some of the major goals of present and future searches in particle
physics. The quadratically divergent contributions to the Higgs
mass and the electroweak precision observables imply different
scales for physics beyond the Standard Model (SM) ($1$ and $10$
TeV respectively). This is the so called little hierarchy problem.
As it is well known the mass of the Higgs boson receives loop
corrections that are quadratic in the loop momenta. The largest
contributions come from the top quark loop, with smaller
corrections coming from loops of the electroweak gauge bosons and
of the Higgs boson itself. Cancellations between the top sector
and other sectors must occur in order to have the Higgs mass
lighter than $200$ GeV as expected from the electroweak precision
test of the Standard Model (SM) which requires a fine-tuning of
one part in 100. As this situation is quite  unnatural
various theories and models have been designed to solve this
problem. For example, in Supersymmetric models the problem of
quadratic Higgs mass divergences is resolved by the introduction
of an opposite-statistic partner for each particle of the SM. The
more recent idea of the Littlest Higgs model (LH) ~\cite{Cohen},
inspired in an old suggestion by Georgi and Pais ~\cite {Georgi},
tries to solve the little hierarchy problem by adding new
particles with masses \textit{O}(TeV) and symmetries which protect
the Higgs mass from those dangerous quadratically divergent
contributions (see~\cite{Schmaltz} and~\cite{review1} for reviews). 
These particles include the Goldstone bosons (GB)
corresponding to a global spontaneous symmetry breaking (SSB) from
the $SU(5)$ to the $SO(5)$ group, a new third generation vector
quark called $T$ and the gauge boson corresponding to an
additional gauge group which contains at least a $SU(2)_R$ and
eventually a new hypercharge $U(1)$. In this case cancellation
occurs between same-statistics particles. However, LH models
typically leave an uncanceled logarithmic mass contributions,
which requires additional new contributions at some high scale to
preserve a small Higgs boson mass.
All of these new states could give rise to a very rich phenomenology,
which could be probed at the CERN Large Hadron Collider 
(LHC)~\cite{Logan,Peskin}.

Nevertheless, it is clear that any viable model has to fulfill  the
basic requirement of reproducing the SM model at low energies. In
particular, from the LH model it is, in principle, possible
 to compute the Higgs low-energy effective potential
and then, by comparing with the SM potential, to obtain their
phenomenological consequences including new restrictions on the
parameter space of the LH model itself. For example, once obtained
the one-loop corrections to the parameters of the standard Higgs
potential, $V=-\mu^{2}HH^{\dag}+\lambda (HH^{\dag})^{2}$;
where $\mu^{2}$ and $\lambda$ denote the well known Higgs mass and
Higgs self-couplings parameters, 
restrictions over the LH parameters space can be obtained by
imposing the condition
 $\mu^2 = \lambda v^2$, where $v$ is the SM vacuum
expectation value ($H=(0,v)/\sqrt{2}$) with $v\simeq 245$ GeV.
The $\mu^{2}$ sign and value are well known~\cite{Cohen,Peskin}, 
and effectively they are the right ones to produce the electroweak 
symmetry breaking, giving a Higgs mass $m_H^2=2 \mu^{2}$. 
However, the full expression for the radiative corrections to 
$\lambda$ has not been analyzed in detail.
In principle both $\mu^2$ and $\lambda$ receive contributions from
fermion, gauge boson and scalar loops, besides others that could
come from the ultraviolet completion of the LH model.
We have previously computed the contributions to the Higgs
effective potential in the LH model coming from the fermion and
gauge boson sectors~\cite{ATP,ATP2}. On the other hand, 
several relations for the threshold
corrections to the $\lambda$ parameter in the presence of a $10$ TeV cut-off,
depending of the UV-completion of the theory, has been reported before
(see, for example~\cite{italianos}).

In this work we continue our program consisting in the computation
of  the relevant terms of the Higgs low-energy effective potential
in the LH model and to analyze their phenomenological
consequences. As has been mentioned before,  
we have started to developed this program in two
previous articles~\cite{ATP,ATP2}. First, we have computed and
analyzed the fermion contributions to the low energy Higgs
effective potential and we have illustrated the kind of
constraints on the possible values of the LH parameters that can
be set by requiring the complete LH Higgs effective potential to
reproduce exactly the SM potential~\cite{ATP}. Second, the effects
of virtual heavy and electroweak gauge bosons present in the LH
model have been included in the analysis~\cite{ATP2}. 
The radiative corrections to $\lambda$, at the one-loop level,
have not been previously computed. First results are presented in the
above two articles. 
We want to note that the computation of $\lambda$ is important 
for several reasons: First, it must be positive, for the low energy
effective action to make sense (otherwise the theory would not
have any vacuum). In addition, from the effective potential above,
one gets the simple formula $m^2_H=2 \lambda v^2$ or,
equivalently, $\mu^2 = \lambda v^2$, being $v$ set by experiment
(for instance from the muon lifetime) to be $v\simeq 245$ GeV.  
Our phenomenological discussion in~\cite{ATP,ATP2} 
have shown that the one-loop effective
potential of the LH model cannot reproduce the SM potential with a
low enough Higgs mass, $m^2_H=2 \lambda v^2=2 \mu^2$, to agree
with the standard expectations. However, there are some
indications suggesting that the effects of including interactions
terms between Goldstone bosons (GB) and the other particles, i.e.
fermions and gauge bosons and/or higher order GB loops could
reduce the Higgs boson mass so that complete compatibility
 with the experimental constraints could be obtained.

The main objective of this work is to compute the effective
potential for the doublet Higgs and the triplet $\phi$, being both
scalar fields of the LH model coming from some of the GB
corresponding to  the global $SU(5)$ to $SO(5)$ global symmetry
breaking of the LH model. Its relevant terms can be read as~\cite{Logan},
\begin{equation}
\label{potef1}
V_{eff}(H,\phi)=-\mu^{2}HH^{\dag}+\lambda (HH^{\dag})^{2}+
\lambda_{\phi^{2}} f^{2} \mbox{tr}(\phi \phi^{\dag})+
i\lambda_{H^2\phi}f(H\phi^{\dag}H^{T}-H^{*}\phi H^{\dag})\,.
\end{equation}
This potential get contributions from
radiative corrections and from effective operators coming from the
ultraviolet completion of the LH model. With this potential we
will study the regions of the LH parameter space giving rise to
the SM electroweak symmetry breaking.
Although radiative corrections from fermion and gauge boson loops 
are discussed in~\cite{ATP,ATP2}, the radiative contributions 
to $\lambda_{\phi^{2}}$ and $\lambda_{H^2\phi}$ have not been computed 
so far. A new constraints over the LH parameter space emerge once
we impose the new relation between coefficients of the effective 
Higgs potential, 
imposed by the diagonalization of the Higgs mass matrix.

This work is organized as follows: In Section 2 we briefly explain
the LH model and set the notation. Section 3 is devoted to the
computation of the  radiative corrections contributions to the
effective potential at one-loop level. Next section is dedicated
to the effective operator contribution. In Section 5 we analyze
the constraints that our computation establishes on the LH
parameters and, finally, in Section 6 we present the conclusions.
The Goldstone bosons couplings to fermions and gauge bosons,
needed for our computations, are listened in the Appendix.

\section{The Littlest Higgs model Lagrangian}

The LH model is based on the assumption that there is a physical
system with a global $SU(5)$ symmetry that is spontaneously broken
to a $SO(5)$ symmetry at a high scale $\Lambda$ through a vacuum
expectation value of order $f$. Thus, 14 Goldstone bosons (GB) are
obtained as a consequence of this breaking. In this work we will
consider two different versions of the LH model. In the first one
 the global $SU(5)$ symmetry is explicitly broken by a gauge group
$[SU(2)\times U(1)]^2$. We refer to this version as \emph{Model
I}~\cite{ATP,ATP2}. In the second one the gauge group  is
$[SU(2)^{2}\times U(1)]$ (\emph{Model II}) ~\cite{ATP,ATP2}. In
both cases some of  the GB become pseudo-GB acquire their masses
through radiative corrections coming from gauge bosons and $t,b$
$T$ fermions loops.

The LH low energy dynamics is then described by a non-linear sigma
model lagrangian plus the appropriate Yukawa terms. The
corresponding lagrangian is given by \cite{Cohen,Logan,Peskin},
\begin{eqnarray}\label{Ltotal}
\textit{L}&=&\textit{L}_{kin}+\textit{L}_{YK}\nonumber\\
&=&\frac{f^2}{8}\mbox{tr} [(D_{\mu}\Sigma)
(D^{\mu}\Sigma)^\dag]-\frac{\lambda_{1}}{2}f
\overline{u}_{R}\epsilon_{mn}\epsilon_{ijk}\Sigma_{im}\Sigma_{jn}\chi_{Lk}-\lambda_{2}
f \overline{U}_{R}U_{L}+\mbox{h.c.}\,,
\end{eqnarray}
where
\begin{equation}
\Sigma=e^{2 i\Pi/f} \Sigma_{0}
\end{equation}
is the GB matrix field. $\Sigma_{0}$ is
\begin{equation}
\Sigma_{0}= \left(%
\begin{array}{ccc}
  0 & 0 & \textbf{1} \\
  0 & 1 & 0 \\
  \textbf{1} & 0 & 0 \\
\end{array}%
\right)\,,
\end{equation}
with  $\textbf{1}$ being the $2 \times 2$ unit matrix. The $\Pi$
matrix can be parametrized as,
\begin{eqnarray}
\Pi & = & \left(%
\begin{array}{ccc}
0& \frac{-i}{\sqrt{2}}H^{\dag} & \phi^{\dag} \\
  \frac{i}{\sqrt{2}}H & 0 & \frac{-i}{\sqrt{2}}H^{*} \\
  \phi & \frac{i}{\sqrt{2}}H^{T} & 0 \\
\end{array}%
\right).
\end{eqnarray}
Here $H=(H^{0},H^{+})$ is the SM Higgs doublet and $\phi$ is the
triplet given by:
\begin{equation}
\phi=\left(%
\begin{array}{cc}
  \phi^{0} & \frac{1}{\sqrt{2}}\phi^{+} \\
\frac{1}{\sqrt{2}}\phi^{+} & \phi^{++}
\end{array}%
\right)\,.
\end{equation}
The covariant derivative $D_{\mu}$ is defined by:
\begin{eqnarray}
\mbox{\emph{Model I}}&&\nonumber\\
D_{\mu}\Sigma & = &
\partial_{\mu}\Sigma-i\sum_{j=1}^{2}g_jW^a_j(Q_j^a\Sigma +\Sigma
Q_j^{aT})- i\sum_{j=1}^{2}g'_jB_j(Y_j\Sigma+\Sigma Y_j^{T})\nonumber\\
\mbox{\emph{Model II}}&&\nonumber\\
D_{\mu}\Sigma & = &
\partial_{\mu}\Sigma-i\sum_{j=1}^{2}g_jW^a_j(Q_j^a\Sigma +\Sigma
Q_j^{aT})- i g'B(Y\Sigma+\Sigma Y^{T})\,,
\end{eqnarray}
where $g$ and $g'$ are the gauge couplings, $W_{j}^{a}$
$(a=1,2,3)$ and $B_{j}\,, B$ are the $SU(2)$ and $U(1)$ gauge
fields, respectively,  $Q_{1ij}^a=\sigma_{ij}^a/2$, for $i,j=1,2$,
$Q_{2ij}^{a}=\sigma_{ij}^{a*}/2$ for $i,j=4,5$ and zero otherwise,
$Y_{1}= diag(-3,-3,2,2,2)/10$, $Y_{2}= diag(-2,-2,-2,3,3)/10$  and
$Y= diag(-1,-1,0,1,1)/2$.

The Yukawa Lagrangian in~(\ref{Ltotal}), $\textit{L}_{YK}$,
describes the interactions between GB and  fermions, more exactly,
the third generations of quarks plus the extra $T$ quark appearing
in the LH model. The indices in $\textit{L}_{YK}$ are defined such
that $m,n=4,5$, $i,j=1,2,3$, and
\begin{eqnarray}
\overline{u}_{R}&=& c \,\overline{t}_{R}+ s  \,\overline{T}_{R}\,,\nonumber\\
\overline{U}_{R}&=&-s \,\overline{t}_{R}+   c \,\overline{T}_{R},
\end{eqnarray}
with:
\begin{eqnarray}
c&=&\cos
\theta=\frac{\lambda_{2}}{\sqrt{\lambda_{1}^{2}+\lambda_{2}^{2}}},\nonumber\\
s&=&\sin \theta =
\frac{\lambda_{1}}{\sqrt{\lambda_{1}^{2}+\lambda_{2}^{2}}}\,,
\end{eqnarray}
and
\begin{equation}
\chi_{L}=\left(%
\begin{array}{c}
  u \\
  b \\
  U \\
\end{array}%
\right)_{L}=\left(%
\begin{array}{c}
  t \\
  b \\
  T \\
\end{array}%
\right)_{L}.
\end{equation}

The $SU(5)$ to $SO(5)$ spontaneous breaking give rise to four
massless gauge bosons  (the SM gauge bosons) and
  four or three massive gauge bosons corresponding to \emph{Model I} or \emph{Model II} respectively.
  In the fermion sector, we obtain one massive $T$ quark
   and two massless quarks, namely the top and the bottom quarks.

In order to compute the gauge bosons loops, the lagrangian $L$
must be supplemented by the standard terms depending only on the
gauge fields. For sake of simplicity we will work in the following
in the Landau gauge. Then these terms can be written symbolically
in the mass eigenstate basis as:
\begin{equation}
\textit{L}_{\Omega}=\frac{1}{2}\Omega^{\mu}((\Box+M_{\Omega}^{2})g_{\mu\nu}
-\partial_{\mu}\partial_{\nu}+2 \tilde{I} \,g_{\mu\nu})\Omega^{\nu}
\end{equation}
where $\Omega$ stands for any of the gauge bosons:
\begin{eqnarray}
\mbox{\emph{Model I}} \hspace{0.5cm} &&
\Omega^{\mu}=({W'}^{\mu a},W^{\mu a},{B'}^{\mu},B^{\mu}) \nonumber\\
\mbox{\emph{Model II}} \hspace{0.5cm} &&
\Omega^{\mu}=({W'}^{\mu a},W^{\mu a},B^{\mu}) \,,
\end{eqnarray}
being the mass matrix eigenstates,
\begin{eqnarray}
\mbox{\emph{Model I}} \hspace{0.5cm} &&
M_{\Omega}=(M_{W'} 1_{3\times 3},0_{3\times 3},M_{B'},0)\nonumber\\
\mbox{\emph{Model II}} \hspace{0.5cm} &&
M_{\Omega}=(M_{W'} 1_{3\times 3},0_{3\times 3},0)\,,
\end{eqnarray}
with $M_{W'}= f \sqrt{g_1^2+g_2^2}/2$ and $M_{B'}= f
\sqrt{g_1^{'2}+g_2^{'2}}/\sqrt{20}$. Finally, $\tilde{I}$ is the
interaction matrix between the gauge bosons and the $H$ and $\phi$
scalars as it is given in the Appendix.

For the quarks, the complete Lagrangian is,
\begin{eqnarray}
\textit{L}_{\chi}= \overline{\chi}_{R}(i
\partials-M+\hat{I})
\chi_{L}+\mbox{h.c.}
\end{eqnarray}
where
$$\chi_{R}=\left(%
\begin{array}{c}
  t \\
  b \\
  T \\
\end{array}%
\right)_{R}\,,$$ $M=$diag$(0,0,m_T)$ with
$m_{T}=f\sqrt{\lambda_{1}^{2}+\lambda_{2}^{2}}$, and $\hat{I}$,
being the scalar-quark interaction matrix, given also in the
Appendix. For more details of the model, including Feynman rules
and also some phenomenological results see, for
example,~\cite{Logan}.

\section{One-loop Effective Potential}

As it is well known the electroweak symmetry breaking in the LH
model is triggered, in principle, by the Higgs potential generated
by one-loop radiative corrections, including both,  fermion and
gauge boson loops. Obviously this potential is invariant under the
electroweak gauge group $SU(2)\times U(1)$. Its relevant terms are given 
in~(\ref{potef1}), being $\mu^{2}$ and $\lambda$ the Higgs mass and the 
Higgs self-couplings parameters respectively. Quartic terms involving $\phi^4$
and $H^2 \phi^2$ are not included since they are not  relevant in
our present computation. The coefficients $\lambda$, $\lambda_{\phi^2}$ and
$\lambda_{H^2\phi}$ appearing in the potential~(\ref{potef1}) receive
contributions  from the tree-level higher order operators coming
from the ultraviolet completion of the LH model (see
Section~\ref{sec:effoperators}) and also from the gauge boson and
fermion radiative corrections as will be discussed in this section.

As described in detail in our previous articles~\cite{ATP,ATP2}, 
we first focused on the effective potential for the $H$
doublet, obtaining the first two terms of the potential,
\begin{equation}
\label{eq:firstV}
V_{eff}(H)=-\mu_{fg}^{2} HH^{\dag}+ \lambda_{fg} (HH^{\dag})^{2}\,,
\end{equation}
where $\mu_{fg}^{2}$ and $\lambda_{fg}$ denote the sum of
fermionic and the gauge boson contributions to $\mu^2$ and
$\lambda$. By imposing that these parameters should reproduce the
SM relation $m_H^2= 2 \lambda v^2= 2\mu^2$, where $m_H$ is the
Higgs mass and $v$ is the vacuum expectation value (vev), we found
that this potential it is not sufficient to find a light Higgs
mass and to satisfy the relation $\mu_{fg}^2=v^2\lambda_{fg}$.
Notice that $v$ is set by the experiment (for instance from the
muon lifetime) to be $v\simeq 245$ GeV and $\mu$ is forced by data
to be at most of order $200$ GeV. However, the inclusion of the
Goldstone boson (GB) sector could channel the situation towards a
complete compatibility with the SM and the experimental
constraints. In this way, the next objective is to obtain the
effective potential for  the $H$ and $\phi$ fields, including the
radiative contributions from fermion and gauge boson loops and the
ones coming from  the effective higher order operators (tree-level
contribution)~\cite{Logan,Cohen,Casas}. In this work we
concentrate on the computation of the fermionic and gauge bosons
contributions to the remaining coefficients of the complete
one-loop effective potential defined in~(\ref{potef1}),
$\lambda_{\phi^2}$ and $\lambda_{H^2\phi}$. For this purpose, we
consider constant GB fields, i.e. $\partial H=\partial\phi=0$.
This assumption makes easier the computation since we have:
\begin{equation}
\label{SV} S_{eff}[H,\phi]=-\int d^{4}x \hspace{0.1cm}
V_{eff}(H,\phi)
\end{equation}
On the other hand the action is quadratic in the fermionic fields.
Therefore, this one-loop contribution is exactly computed.

We split the calculation in two parts: the first one is dedicated to the fermion sector,
 and the second one to the gauge boson sector.
Details on how the effective action is computed, by using standard
techniques (see for instance~\cite{book}), are given
in~\cite{ATP,ATP2}. In the following we summarize just the main
steps needed for the calculation.

\subsection{Fermionic contribution}

Following the idea in~\cite{ATP}, the fermionic part of the
effective action can be expanded as:
\begin{eqnarray} \label{expfer}
S_{eff}^{f}[H,\phi]&\simeq&-i\mbox{Tr}\log(1+G \tilde{I}_{f})=-i\mbox{Tr}\sum_{k=1}^{\infty}\frac{(-1)^{k+1}}{k}(G\tilde{I}_{f})^{k},
\end{eqnarray}
where we have neglected a constant, irrelevant for the computation
of the effective action. The fermion propagator, $G^{ab}(x,y)$, is given by:
\begin{equation}
G^{ab}(x,y) \equiv \int d\tilde{k}e^{-ik(x-y)}(\kslash -m_{f})^{-1}_{ab}  \hspace{2cm} a,b\equiv t,b,T,
\end{equation}
where $d\tilde k \equiv d^4 k/(2 \pi)^4$,
and the interaction operators are:
\begin{equation}
\hat{\tilde{I}}_{f}^{ab}(x,y)=(\tilde{I}_{f1}+\tilde{I}_{f2}+\tilde{I}_{f3}+\tilde{I}_{f4})\delta(x-y)\delta^{ab}.
\end{equation}
Here the subindex indicates the number of GB interacting with two fermions.

\begin{figure}
\begin{center}
\epsfig{file=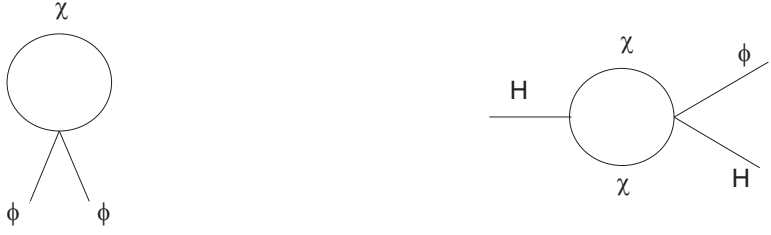,scale=0.4}
\end{center}\vspace*{-0.6cm}
\caption{Fermionic one-loop diagrams contributing to the $\lambda_{\phi^2}$ and $\lambda_{H^2\phi}$, with $\chi=t$,$b$,$T$. All possible combinations of these particles appear in the loops.}
\label{loopfer}
\end{figure}

In order to obtain the fermionic contribution to the $\lambda_{\phi^2}$
and $\lambda_{H^2\phi}$ we only need consider the terms $k=1$ and $k=2$ in the expansion (\ref{expfer}), respectively. The generic one-loop diagrams are shown in Fig.\ref{loopfer}. For $k=1$ we get,
\begin{equation}
S_{f}^{(1)}[H,\phi]=-i\mbox{Tr}[G^{a}(\tilde{I}_{f2}^{aa}+\tilde{I}_{f4}^{aa})]\,.
\end{equation}
For the case $k=2$ one obtains,
\begin{equation}
S_{f}^{(2)}[H,\phi]=\frac{i}{2}\mbox{Tr}[2G^{a}\tilde{I}_{f1}^{ab}G^{b}\tilde{I}_{f2}^{ba}+G^{a}\tilde{I}_{f2}^{ab}G^{b}\tilde{I}_{f2}^{ba}+2G^{a}\tilde{I}_{f1}^{ab}G^{b}\tilde{I}_{f3}^{ba}].
\end{equation}

By using well known methods, and after some work in which the
divergent integrals that
 emerge are regularized by using an ultraviolet cutoff $\Lambda$, we obtain the different contributions to the couplings. The Fermionic
  contributions are:
\begin{eqnarray}\label{lphif}
\lambda_{\phi^2f}&=&\frac{8N_{c}}{(4\pi f)^{2}}(\lambda_{t}^{2}+\lambda_{T}^{2})\left(\Lambda^{2}-m_{T}^{2}
\log\left(\frac{\Lambda^{2}}{m_{T}^{2}}+1\right)\right),
\end{eqnarray}
and
\begin{eqnarray}\label{lh2phif}
\lambda_{H^{2}\phi\,f}&=&-\frac{4N_{c}}{(4\pi f)^{2}}\left[(\lambda_{t}^{2}+\lambda_{T}^{2})\Lambda^{2}-\lambda_{T}^{2}m_{T}^{2} \log\left(\frac{\Lambda^{2}}{m_{T}^{2}}+1\right)\right]\,,
\end{eqnarray}
where $N_c$ is the number of colors and,  $\lambda_t$ and
$\lambda_T$ are, respectively, the SM top Yukawa coupling and the
heavy top Yukawa coupling, given by:
\begin{equation}
\lambda_{t}=
\frac{\lambda_{1}\lambda_{2}}{\sqrt{\lambda_{1}^{2}+\lambda_{2}^{2}}}\,,\,\,\,\,\,\,
\lambda_{T}=\frac{\lambda_{1}^{2}}{\sqrt{\lambda_{1}^{2}+\lambda_{2}^{2}}}\,.
\end{equation}

For the purpose of illustration and the final discussion of this
paper, we also summarize here the fermionic contribution to the other two
parameters of the Higgs potential, $\mu^{2}$ and $\lambda$,
as have been obtained in~\cite{ATP}:
\begin{equation}\label{muf}
\mu^{2}_f= N_{c} \frac{m_{T}^{2} \lambda_{t}^{2}}{4 \pi^{2}}
\log\left(1+\frac{\Lambda^{2}}{m_{T}^{2}}\right),
\end{equation}
and
\begin{eqnarray}\label{lf}
\lambda_f&=&\frac{N_{c}}{(4
\pi)^{2}}\left[2(\lambda_{t}^{2}+\lambda_{T}^{2})
\frac{\Lambda^{2}}{f^{2}}-\log\left(1+\frac{\Lambda^{2}}{m_{T}^{2}}\right)
\left(-\frac{2m_{T}^{2}}{f^{2}}\left(\frac{5}{3}
\lambda_{t}^{2}+\lambda_T^{2}\right)+4\lambda_{t}^{4}
+4(\lambda_{T}^{2}+
\lambda_{t}^{2})^{2}\right)\right.\nonumber\\
&-&\left.4\lambda_{T}^{2}\frac{1}{1+\frac{m_{T}^{2}}{\Lambda^{2}}}\left
(\frac{m_{T}^{2}}{f^{2}}-2\lambda_{t}^{2} -\lambda_{T}^{2}\right)
-4\lambda_{t}^{4}
\log\left(\frac{\Lambda^{2}}{m^{2}}\right)\right].
\end{eqnarray}

Observe that the $\lambda$'s parameters, $\lambda_f,
\lambda_{\phi^2f}$ and $\lambda_{H^{2}\phi\,f}$, are quadratically
divergent. This is due to the lack of any symmetry  protecting
them, unlike the $\mu$ parameter which is protected by a $SU(3)$
global symmetry. This will be the case for the  gauge sector too,
as it will be seen in the following.

\subsection{Bosonic contribution}

Here we concentrate in the gauge boson contributions at the
one-loop level. We use the Landau gauge so that we do not have to
consider any ghost field at this level. In this case the effective
action can be expand as:
\begin{equation}\label{expgb}
S_{eff}^{g}[H,\phi]=\frac{i}{2}\mbox{Tr}\log(1+G \tilde{I}_{g})
=\frac{i}{2}\mbox{Tr}
\sum_{k=1}^{\infty}\frac{(-1)^{k+1}}{k}(2G \tilde{I}_{g})^{k},
\end{equation}
where $G$ is the gauge boson propagator given by (Landau gauge):
\begin{equation}
G^{ab}_{\mu\nu}(x,y) \equiv \int d\tilde{k}\,
\frac{e^{-ik(x-y)}}{k^{2}-M_{g}^{2}}
{\left(-g_{\mu\nu}+\frac{1}{k^{2}}k_{\mu}k_{\nu}\right)}_{ab}\,,
\hspace{1.5cm} a,b={W'}^{a},W^{a},B',B.
\end{equation}
and the interaction operators are:
\begin{equation}
\hat{\tilde{I}}_{g}^{ab}(x,y)=(\tilde{I}_{g2}+\tilde{I}_{g3}+\tilde{I}_{g4})\delta(x-y)\delta^{ab}.
\end{equation}

\begin{figure}[t!]
\begin{center}
\epsfig{file=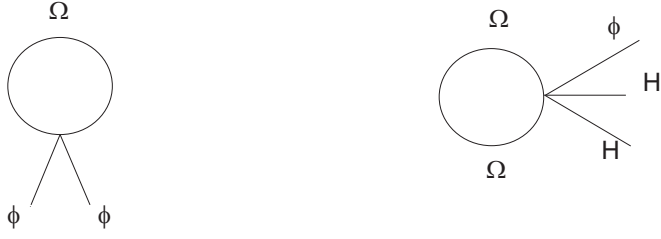,scale=0.4}
\end{center}\vspace*{-0.6cm}
\caption{One-loop gauge boson diagrams contributing to the $\lambda_{\phi^2}$ and $\lambda_{H^2\phi}$, where $\Omega$ represents to the gauge bosons particles, $W^{'1,2,3}$, $W^{1,2,3}$, $B'$ and $B$. All possible combinations of these particles appear in the loops.}
\label{loopbg}
\end{figure}
The generic diagrams for this computation are given in
Fig.\ref{loopbg}. In this case, we only need to consider the term
$k=1$ to obtain the two parameters $\lambda_{\phi^2}$ and
$\lambda_{H^2\phi}$. Then, we get:
\begin{equation}
S_{g}^{(1)}[H,\phi]=i\mbox{Tr}[G(\tilde{I}_{g2}+\tilde{I}_{g3}+\tilde{I}_{g4})]
\end{equation}

As it was mentioned above we consider in our analysis two
different models: The original LH with two $U(1)$ groups (\emph{Model I}) and the other one with
  just one $U(1)$ group (\emph{Model II}). As there is no mixing between
  the $SU(2)$ and $U(1)$ groups, the only difference among these two models
occurs in the $U(1)$ sector.

\subsubsection{\emph{Model I}}

The contributions from gauge boson sector to the
$\lambda_{\phi^2}$ and $\lambda_{H^2\phi}$ parameters are given
by:
\begin{eqnarray}
\label{lphigI}
\lambda^{I}_{\phi^{2} g}&=&\frac{3}{4(4 \pi f)^2}\left[\frac{g^{2}}{c_{\psi}^{2}s_{\psi}^{2}}\Lambda^{2}-g^{2}M_{W'}^{2}\log\left(\frac{\Lambda^{2}}{M_{W'}^{2}}+1\right)\left(\frac{(s_{\psi}^{2}-c_{\psi}^{2})^{2}}{c_{\psi}^{2}s_{\psi}^{2}}-4 \right)\right.\nonumber\\
&+&\left.\frac{g^{'2}}{c_{\psi'}^{2}s_{\psi'}^{2}}\Lambda^{2}-g^{'2}M_{B'}^{2}\log\left(\frac{\Lambda^{2}}{M_{B'}^{2}}+1\right)\frac{(s_{\psi'}^{2}-c_{\psi'}^{2})^{2}}{c_{\psi'}^{2}s_{\psi'}^{2}}\right]\,,
\end{eqnarray}
\begin{eqnarray}\label{lh2phigI}
\lambda^{I}_{H^2\phi g}&=&\frac{3}{8(4\pi f)^{2}}\left[g^{2}\frac{s_{\psi}^{2}-c_{\psi}^{2}}{c_{\psi}^{2}s_{\psi}^{2}}\left(\Lambda^{2}-M_{W'}^{2}\log\left(\frac{\Lambda^{2}}{M_{W'}^{2}}+1\right)\right)\right.\nonumber\\
&+&\left.g^{'2}\frac{s_{\psi'}^{2}-c_{\psi'}^{2}}{c_{\psi'}^{2}s_{\psi'}^{2}}\left(\Lambda^{2}-M_{B'}^{2}\log\left(\frac{\Lambda^{2}}{M_{B'}^{2}}+1\right)\right)\right]\,.
\end{eqnarray}

Note that this last parameter only receives contributions from the
heavy gauge boson sector.

For the sake of completeness and our  phenomenological discussion
we also list here the results for the gauge boson contributions to
$\mu^{2}$ and $\lambda$, as obtained in~\cite{ATP2}:
\begin{eqnarray}\label{muM1}
\mu^{2\,I}_{g}&=&-\frac{3}{64\pi^{2}}\left[3g^{2}M_{W'}^{2}
\log\left(1+\frac{\Lambda^{2}}{M_{W'}^{2}}\right)+g^{'2}M_{B'}^{2}\log\left(1+\frac{\Lambda^{2}}{M_{B'}^{2}}
\right)\right],
\end{eqnarray}
\begin{eqnarray}\label{lM1}
\lambda^{I}_{g}&=&-\frac{3}{(16\pi f)^{2}}\left[-\left(\frac{g^{2}}{c_{\psi}^{2}s_{\psi}^{2}}+
\frac{g^{'2}}{c_{\psi}^{'2}s_{\psi}^{'2}}\right)\Lambda^{2}\right.\nonumber\\
&+&\left.g^{2}M_{W'}^{2}
\log\left(1+\frac{\Lambda^{2}}{M_{W'}^{2}}\right)\left(4+\frac{1}{c_{\psi}^{2}s_{\psi}^{2}}
+2g^{'2}\frac{(c_{\psi}^{2}s_{\psi}^{'2}+s_{\psi}^{2}c_{\psi}^{'2})^{2}}{c_{\psi}^{2}
s_{\psi}^{2}c_{\psi}^{'2}s_{\psi}^{'2}}\frac{f^{2}}{M_{W'}^{2}-M_{B'}^{2}}\right)
\right.\nonumber \\
&+&\left.g^{'2}M_{B'}^{2} \log\left(1+\frac{\Lambda^{2}}{M_{B'}^{2}}\right)
\left(\frac{4}{3}+\frac{1}{c_{\psi}^{'2}s_{\psi}^{'2}}
+2g^{2}\frac{(c_{\psi}^{2}s_{\psi}^{'2}+s_{\psi}^{2}c_{\psi}^{'2})^{2}}{c_{\psi}^{2}
s_{\psi}^{2}c_{\psi}^{'2}s_{\psi}^{'2}}\frac{f^{2}}{M_{B'}^{2}-M_{W'}^{2}}\right)
\right.\nonumber \\
&+&\left.f^{2}\log\left(1+\frac{\Lambda^{2}}{M_{W'}^{2}}\right)\left(3g^{4}+
2(3g^{2}+g^{'2})g^{2}\frac{(s_{\psi}^{2}-c_{\psi}^{2})^{2}}{c_{\psi}^{2}s_{\psi}^{2}}\right)\right.\nonumber\\
&+&\left.f^{2}\log\left(1+\frac{\Lambda^{2}}{M_{B'}^{2}}\right)\left(g^{'4}+2(g^{2}+
g^{'2})g^{'2}\frac{(s_{\psi}^{'2}-c_{\psi}^{'2})^{2}}{c_{\psi}^{'2}s_{\psi}^{'2}}
\right)\right.\nonumber \\
&+&\left.f^{2}\log\left(\frac{\Lambda^{2}}{m^{2}}\right)\left(3g^{4}+g^{'4}+
8g^{2}g^{'2}\right)-3f^{2}\frac{g^{4}}{1-\frac{M_{W'}^{2}}{\Lambda^{2}}}-
f^{2}\frac{g^{'4}}{1-\frac{M_{B'}^{2}}{\Lambda^{2}}}\right]\,.
\end{eqnarray}
where
\begin{eqnarray}
g &\equiv&=\frac{g_{1}g_{2}}{\sqrt{g_{1}^{2}+g_{2}^{2}}}, \hspace{0.25cm} s_{\psi} = \sin \psi = \frac{g_1}{\sqrt{g_1^2+g_2^2}}, \hspace{0.25cm}
c_{\psi} = \cos \psi =\frac{g_2}{\sqrt{g_1^2+g_2^2}}\,
\end{eqnarray}
and
\begin{eqnarray}
g'&\equiv&\frac{g'_{1}g'_{2}}{\sqrt{g_{1}^{'2}+g_{2}^{'2}}},\hspace{0.25cm}
s'_{\psi}=\sin \psi' = \frac{g'_{1}}{\sqrt{{g'}_{1}^{\,2}+{g'}_{2}^{\,2}}},\hspace{0.25cm}
c'_{\psi}=\cos\psi'=\frac{{g'}_{2}}{\sqrt{{g'}_{1}^{\,2}+{g'}_{2}^{\,2}}}\,.
\end{eqnarray}

\subsubsection{\emph{Model II}}

The corresponding results for this model are:
\begin{eqnarray} \label{lphigII}
\lambda^{II}_{\phi^{2} g}&=&\frac{3}{64 \pi^{2}f^2}\left[\frac{g^{2}}{c_{\psi}^{2}s_{\psi}^{2}}\Lambda^{2}-g^{2}M_{W'}^{2}\log\left(\frac{\Lambda^{2}}{M_{W'}^{2}}+1\right)\left(\frac{(s_{\psi}^{2}-c_{\phi}^{2})^{2}}{c_{\psi}^{2}s_{\psi}^{2}}-4 \right) \right]+\frac{3g^{'2}}{(4\pi f)^{2}}\Lambda^{2},\nonumber\\
\end{eqnarray}
\begin{eqnarray}\label{lh2phigII}
\lambda^{II}_{H^{2}\phi g}&=&\frac{3g^{2}}{8(4f\pi)^{2}}\frac{s_{\psi}^{2}-c_{\psi}^{2}}{c_{\psi}^{2}s_{\psi}^{2}}\left(\Lambda^{2}-M_{W'}^{2}\log\left(\frac{\Lambda^{2}}{M_{W'}^{2}}+1\right)\right).
\end{eqnarray}

As it was expected, the $U(1)$ sector does not have any influence
on $\lambda^{II}_{H^2\phi g}$.

In addition the $\mu^{2\,II}_{g}$ and $\lambda^{II}_{g}$
parameters are given by ~\cite{ATP2}:
\begin{equation}\label{muM2}
\mu^{2\,II}_{g}=-\frac{3}{64\pi^{2}}\left(3g^{2}M_{W'}^{2}
\log\left(1+\frac{\Lambda^{2}}{M_{W'}^{2}}\right)+g^{'2} \Lambda^{2}\right),
\end{equation}
and
\begin{eqnarray}\label{lM2}
\lambda^{II}_{g}&=&-\frac{3}{(16 \pi f)^{2}}
\left[-\frac{g^{2}}{c_{\psi}^{2}s_{\psi}^{2}}\Lambda^{2}+
\frac{4}{3}{g'}^{2}\Lambda^{2}+ g^{2}M_{W'}^{2}
\log\left(\frac{\Lambda^{2}}{M_{W'}^{2}}+1\right)
\left(4+\frac{1}{c_{\psi}^{2}s_{\psi}^{2}}\right) \right. \nonumber \\
&+&\left.f^{2}\log \left(1+\frac{\Lambda^{2}}{M_{W'}^{2}}\right)
\left(3g^{4}+2(3g^{2}+{g'}^{2})g^{2}\frac{(s_{\psi}^{2}-c_{\psi}^{2})^{2}}
{s_{\psi}^{2}c_{\psi}^{2}}\right)\right. \nonumber\\
&+&\left.f^{2}\log\left(\frac{\Lambda^{2}}{m^{2}}\right)
(3g^{4}+{g'}^{4}+8g^{2}{g'}^{2})
-3f^{2}\frac{g^{4}}{1-\frac{M_{W'}^{2}}{\Lambda^{2}}}\right].
\end{eqnarray}

With these results, the radiative contributions at one-loop level
to the Higgs potential parameters are completed.

\section{Effective Operators}
\label{sec:effoperators}

As discussed above, the Higgs potential  gets gauge boson and
fermion one-loop contributions in the LH model. In addition, the
potential coefficients also receive contributions from additional
operators coming from the ultraviolet completion of the LH model.
Thus these operators must be consistent with the symmetry of the
theory ~\cite{Cohen,Logan,Casas}. At the lowest order they can be
parameterized by two unknown coefficients $a$ and $a'$ $\sim O(1)$. 
The form of these effective operators are, for the fermion sector~\cite{Logan},
\begin{equation}
\textit{O}_{f}=-a'\frac{1}{4}\lambda_{1}^{2}f^{4}\epsilon^{wx}\epsilon_{yz}\epsilon^{ijk}\epsilon_{kmn}\Sigma_{iw}\Sigma_{jx}\Sigma^{*my}\Sigma^{*nz}\,,
\end{equation}
where $i,j,k,m,n$ run over 1,2,3 and $w,x,y,z$ run over 4,5 and
for
 the gauge sector (\emph{Model I}),
\begin{eqnarray}
\textit{O}_{gb}=\frac{1}{2}af^{4}\left\{g_{j}^{2}\sum_{a=1}^{3}\mbox{Tr}\left[(Q_{j}^{a}\Sigma)(Q_{j}^{a}\Sigma)^{*}\right]+g_{j}^{'2}\mbox{Tr}\left[(Y_{j}\Sigma)(Y_{j}\Sigma)^{*}\right]\right\}\,,
\end{eqnarray}
with $j=1,2$ and $Q_{j}^{a}$ and  $Y_{j}$ being
the generators of the $SU(2)_{j}$ and $U(1)_{j}$ groups, respectively.

In the case of the \emph{Model II}:
\begin{eqnarray}
\textit{O}_{gb}=\frac{1}{2}cf^{4}\left\{g_{j}^{2}\sum_{a=1}^{3}\mbox{Tr}\left[(Q_{j}^{a}\Sigma)(Q_{j}^{a}\Sigma)^{*}\right]+g^{'2}\mbox{Tr}\left[(Y\Sigma)(Y\Sigma)^{*}\right]\right\}\,,
\end{eqnarray}
where $j=1,2$ and $Y$ is the generator of the unique $U(1)$ group.

By expanding the GB field matrix $\Sigma$ in these effective
operators, we obtain their different contributions to the
coefficients of the effective potential (\ref{potef1}):
\begin{center}
\begin{tabular}{lll} \label{Table}
parameters & \emph{Model I} & \emph{Model II} \\
$\lambda_{\rm EO}$ & $\frac{a}{8}\left(\frac{g^2}{s_{\psi}^2c_{\psi}^2}+\frac{g^{'2}}{s_{\psi}^{'2}c_{\psi}^{'2}}\right)+2a'(\lambda_{t}^2+\lambda_{T}^2)$ & $\frac{a}{8}\left(\frac{g^2}{s_{\psi}^2c_{\psi}^2}\right)-\frac{a}{3}g^{'2}+2a'(\lambda_{t}^2+\lambda_{T}^2)$\\
${\lambda_{\phi^2}}_{\rm EO}$ &  $\frac{a}{2}\left(\frac{g^2}{s_{\psi}^2c_{\psi}^2}+\frac{g^{'2}}{s_{\psi}^{'2}c_{\psi}^{'2}}\right)+8a'(\lambda_{t}^2+\lambda_{T}^2)$ & $\frac{a}{2}\left(\frac{g^2}{s_{\psi}^2c_{\psi}^2}\right)+4a{g^{'2}}+8a'(\lambda_{t}^2+\lambda_{T}^2)$\\
${\lambda_{H^2\phi}}_{\rm EO}$ & $\frac{a}{4}\left(g^2\frac{c_{\psi}^2-s_{\psi}^2}{s_{\psi}^2c_{\psi}^2}+g^{'2}\frac{c_{\psi}^{'2}-s_{\psi}^{'2}}{s_{\psi}^{'2}c_{\psi}^{'2}}\right)+4a'(\lambda_{t}^2+\lambda_{T}^2)$ & $\frac{a}{4}g^2\frac{c_{\psi}^2-s_{\psi}^2}{s_{\psi}^2c_{\psi}^2}+4a'(\lambda_{t}^2+\lambda_{T}^2)$\\
$\mu^{2}_{\rm EO}$ & $0$ & $a f^2 g^{'2}$
\end{tabular}
\end{center}

To summarize, the complete results for these parameters is the sum
of the contributions coming from the effective
operators, as given above, and the radiative contributions coming
from the fermion and gauge boson sector, which were given in
Section 3.

\section{Numerical Results and Phenomenological Discussion}

In this section we discuss about the constraints on the possible
values of the LH parameters. In our previous works we focused on
the analysis of the constraints on the LH parameters by
considering the effective potential only for the LH
doublet~(\ref{eq:firstV})~\cite{ATP,ATP2}. Our computation
included the effect of virtual heavy quarks $t, b$ and $T$,
together with the heavy and electroweak gauge bosons $W^{'}$, $W$,
$B'$ and $B$, present in the LH model. By imposing that the
potential~(\ref{eq:firstV}) has a minimum whenever $\mu^2=\lambda
v^2$ ($v=246$ GeV), we found that the obtained values for the
$\mu$ parameter were too high to be compatible with the expected
Higgs mass, which should not be larger than about 200 GeV
according to the electroweak precision data.

It is clear that a similar analysis should be done if we consider
the complete effective Higgs potential as given in~(\ref{potef1}).
In this case, by diagonalizing the Higgs mass matrix, the Higgs
mass  is given to the leading order by $m_{H}^2\simeq
2(\lambda-\lambda_{H^2\phi}^2/\lambda_{\phi^2})v^2 =2 \mu^2$
~\cite{Logan}. Therefore, the LH parameters must satisfy the
condition:
\begin{eqnarray}\label{cond}
v^2=\frac{\mu^{2}}{(\lambda-\lambda_{H^2\phi}^2/\lambda_{\phi^2})}\,.
\end{eqnarray}

In the following we will discuss about the constraints that the
condition (\ref{cond}) imposes on the LH parameters space. In this
way, we should also take into account other constraints imposed by
requiring the consistency of the LH models with the electroweak
precision data. There exist several studies of the corrections to
electroweak precision observables in the Little Higgs models,
exploring whether there are regions of the parameter space in
which the model is consistent with
data~\cite{Logan,Peskin,Csaki1,Csaki2,Schmaltz,review1,LoganP,EWPO1,recentpheno}.
In the  \emph{Model I} with a gauge group $SU(2)\times SU(2)
\times U(1) \times U(1)$ one has a multiplet of heavy $SU(2)$
gauge bosons and a heavy $U(1)$ gauge boson. The last one leads to
large electroweak corrections and some problems with the direct
observational bounds on the $Z'$ boson from
Tevatron~\cite{Csaki1,Csaki2}. Then, a very strong bound on the
symmetry breaking scale $f$, $f>4$ TeV at $95 \%$ C.L, is
found~\cite{Csaki1}. However, it is known that this bound is
lowered to $1-2$ TeV for some region of the parameter
space~\cite{Csaki2} by gauging only $SU(2)\times SU(2) \times
U(1)$ (\emph{Model II}). Thus in the following, we focus in the LH
version called \emph{Model II}.

In order to avoid small values of the $W'$ mass and a very strong
coupling constant  we will set $M_{W'}
>0.6$ TeV and $g_{R}\leq 3$ in our numerical discussion. We have found that for very small or
very large values of the gauge group mixing angles, $\mu_{fg}^2$
is not positive and the SSB does not occur. However, due to the
dependence of heavy gauge coupling constants on the mixing angles,
\begin{equation}
g_{R}^{2} \equiv \frac{1}{2}(g_{1}^{2}s_{\psi}^{2}+g_{2}^{2}c_{\psi}^{2}),
\end{equation}
it is found that $c_{\psi}< 0.1$ or $c_{\psi}\sim1$ imply a very
strong gauge coupling.  Accordingly we will work with $0.1<
c{_\psi}<0.9$, and then we will ensure that $\mu^2$ has the right
sing to generate a SSB. Besides, taking into account the
restrictions on the parameters given in~\cite{ATP2}, we also set
the following ranges: $0.5<\lambda_{T}<2$, $0.8$ TeV $<f<1$ TeV
and $10$ TeV $<\Lambda<12$ TeV. The condition $\lambda_T \gsim$
0.5 is established setting the bounds on the couplings $\lambda_1,
\lambda_2 \geq m_t/v$ or $\lambda_1 \lambda_2 \geq 2 (m_t/v)^2$
from the top mass~\cite{Logan}. The condition  $m_T\lsim$ 2.5 TeV
is required in order to avoid a large amount of fine-tuning in the
Higgs potential~\cite{Cohen,Peskin}. Then, since $m_{T}$ grows
linearly with  $f$, $f$ should be lesser than about one
TeV~\cite{ATP}. Finally, $\Lambda$ is restricted by the standard
condition $\Lambda \sim 4 \pi f$~\cite{relationlambdaf}. On the
other hand the  $a$ and $a'$ parameters are expected to be $O(1)$
~\cite{Logan,Cohen,Casas}. Values of the symmetry breaking scale
$f$ around $4$ TeV are also allowed by the  electroweak precision
observables~\cite{Csaki1,Csaki2}. However, this value of $f$
implies that $m_T$ is always greater than 5.7 TeV, when $\lambda_T
> 0.5$. A fine-tuning of $0.8\%$ is estimated for a Higgs mass of
$200$ GeV~\cite{Csaki1}. Besides, one gets $M_{W'} > 2.6$ TeV. In
addition, we have found that for $f=4$ TeV, the allowed region for
the LH parameters space, satisfying the condition imposed by the
minimum of the Higgs potential, is smaller ~\cite{ATP,ATP2}. In
fact, values around $1-3$ TeV are the preferred ones for our
selected choices of the LH parameters.

By considering the LH parameters bounded as described above and
imposing the condition (\ref{cond}), we analyze the constraints 
on the different LH parameters with the inclusion of the effects 
of both contributions, the radiative ones (for the \emph{Model II})
and the effective operators. 
We found that the minimum $\mu$ value  is $\mu=0.39$ TeV;
for $f=0.8$ TeV, $\Lambda=10$ TeV, $\lambda_{T}=0.72$,
$c_{\psi}=0.34$ and $a=a'=0$, and  the maximum is $\mu=0.761$ TeV;
for $f=0.95$ TeV, $\Lambda=11.5$ TeV, $\lambda_{T}=1.8$,
$c_{\psi}=0.71$, $a=1$ and $a'=0.3$. Clearly, the minimum value 
corresponds with the case of considering the radiative corrections alone.
In this case, small values of $\cos {\psi}$ are preferred. This is due to the
fact that the  fermionic contribution to the $\mu$ parameter is
very large for higher energies, while $\lambda$ does not change
strongly with $f$ and $\Lambda$. In this way, to satisfy the
condition (\ref{cond}), small values of $\cos {\psi}$ are needed
in order to reduce $\mu$.
As it was expected, large  $\phi$ mass, $M_{\phi}=4.1$ TeV, is found 
for values of the parameters
corresponding to the minimum value of $\mu$.

\begin{figure}[t!]
\begin{center}
\begin{tabular}{cccc}\hspace{-1cm}
\epsfig{file=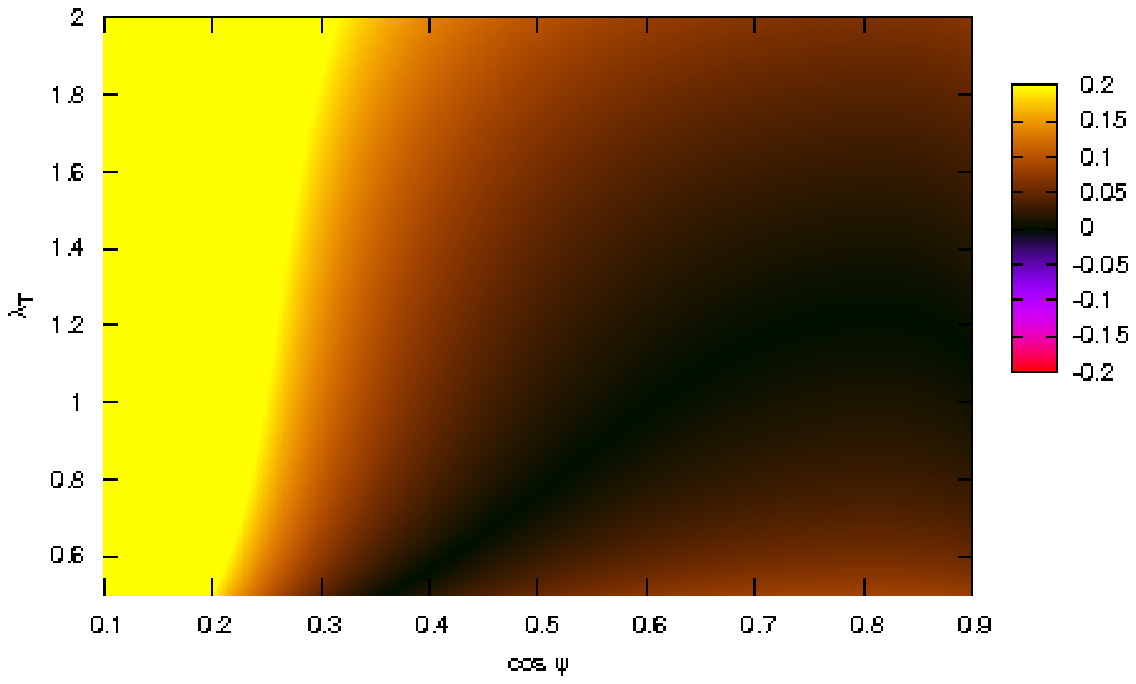,scale=0.4} & \hspace{-1.5cm}\epsfig{file=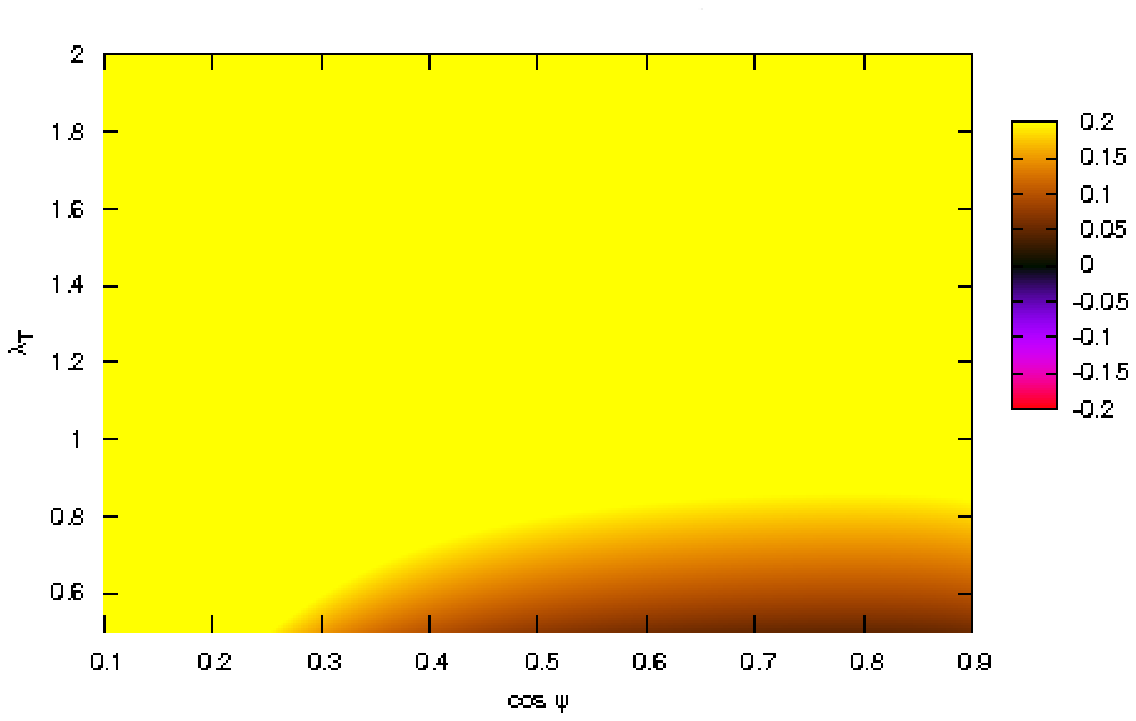,scale=0.4}
& \hspace{-1.5cm} \epsfig{file=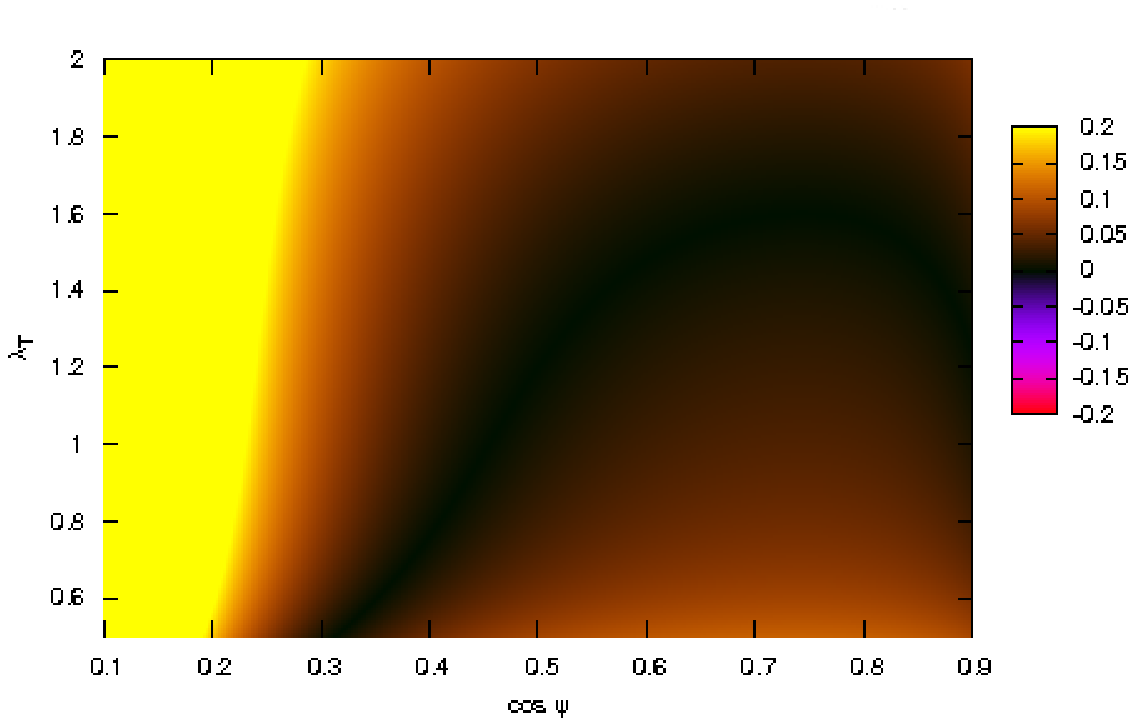,scale=0.4} & \hspace{-1.5cm} \epsfig{file=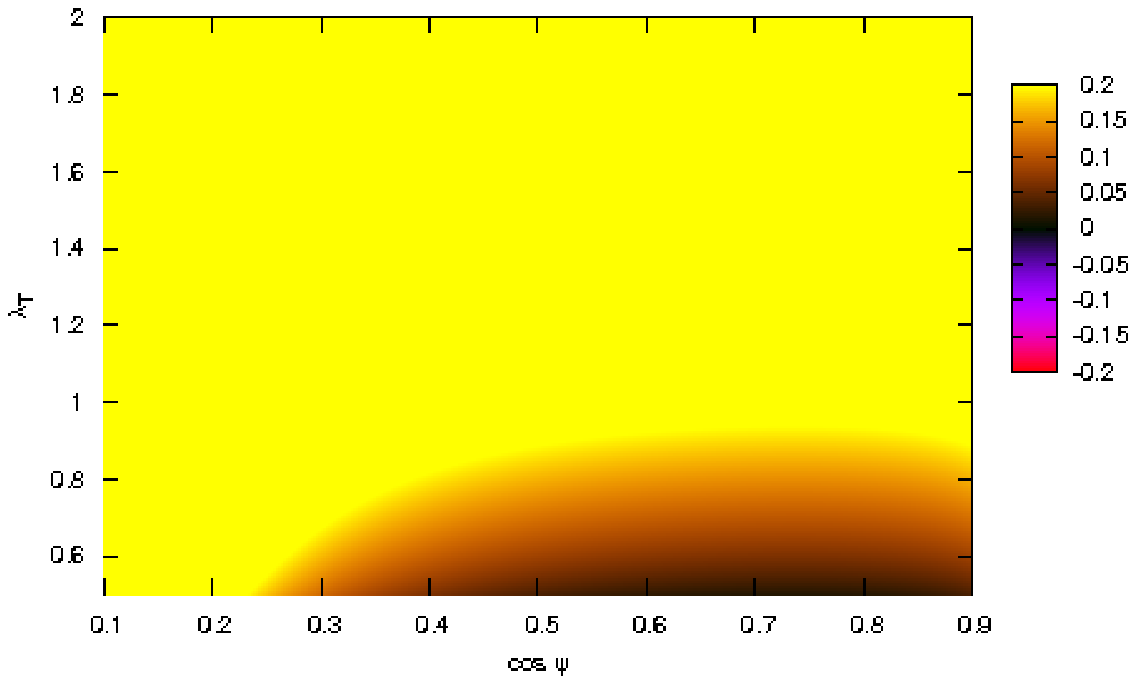,scale=0.4}\\\hspace{-1cm}
\epsfig{file=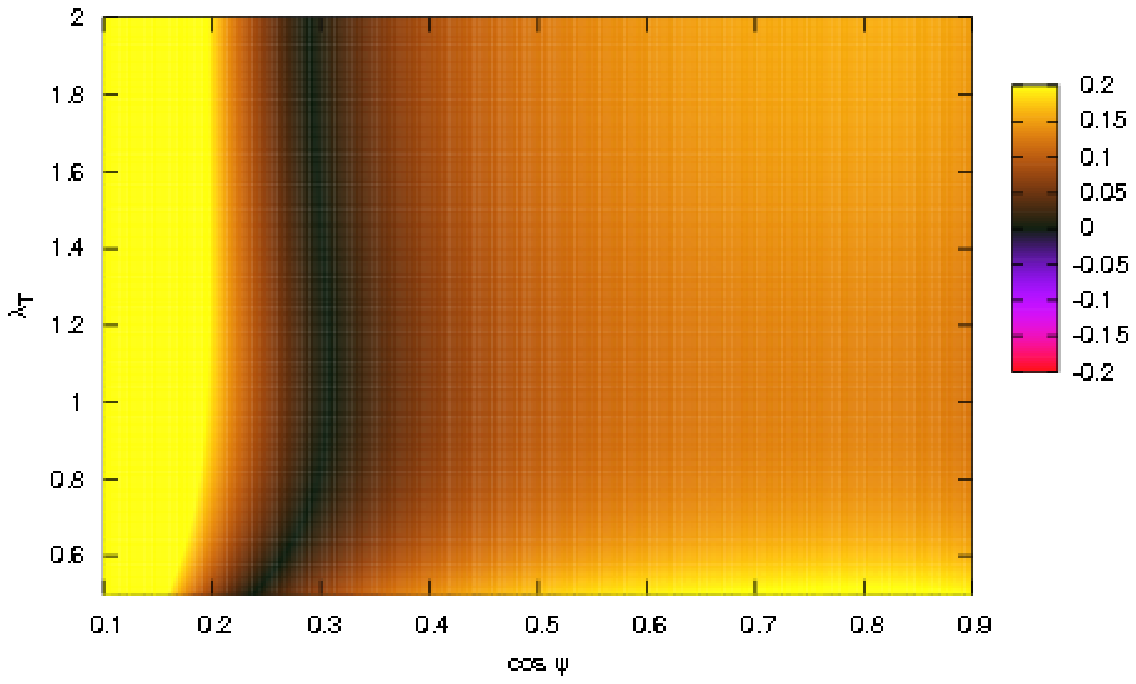,scale=0.4} & \hspace{-1.5cm} \epsfig{file=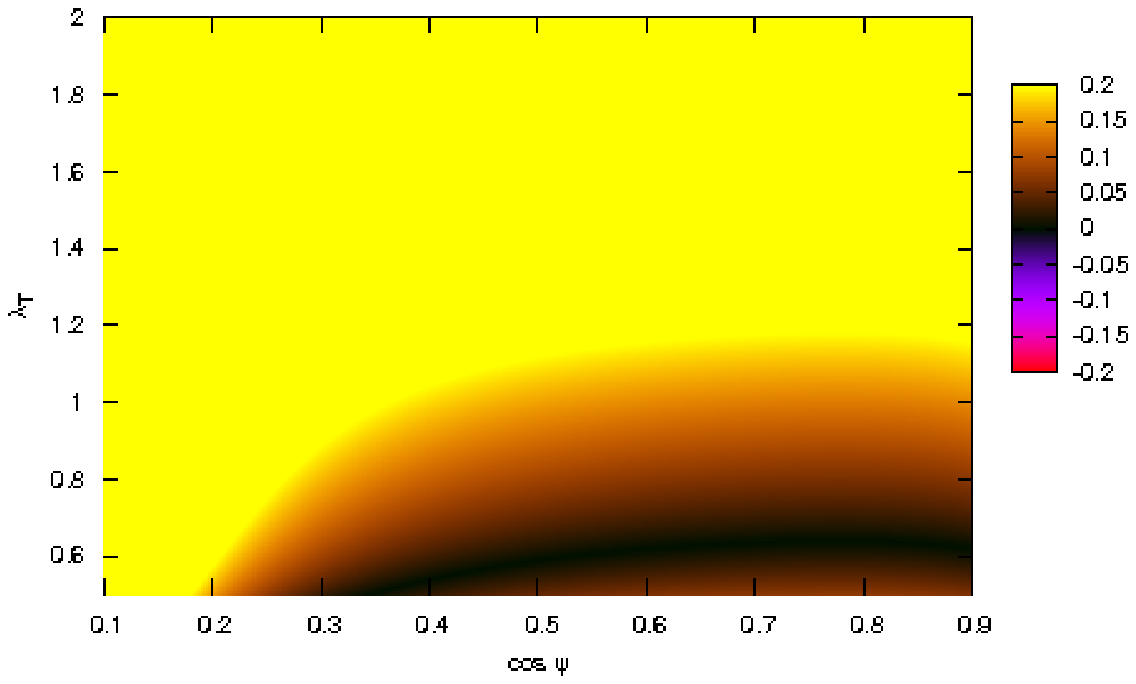,scale=0.4}
& \hspace{-1.5cm} \epsfig{file=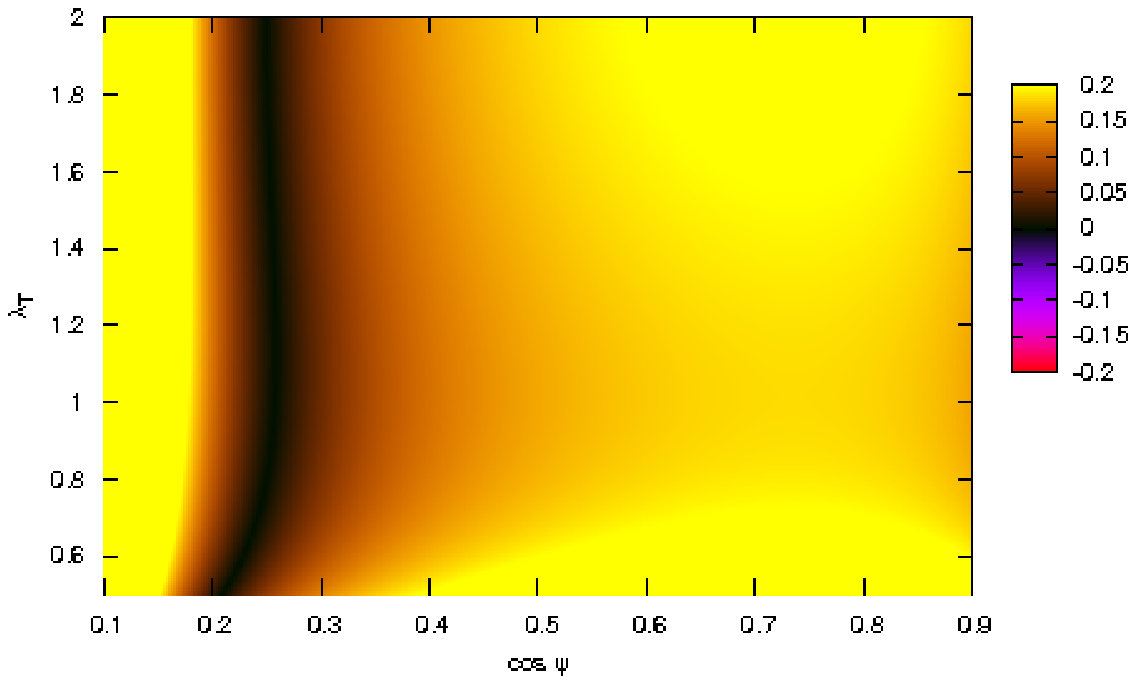,scale=0.4} & \hspace{-1.5cm} \epsfig{file=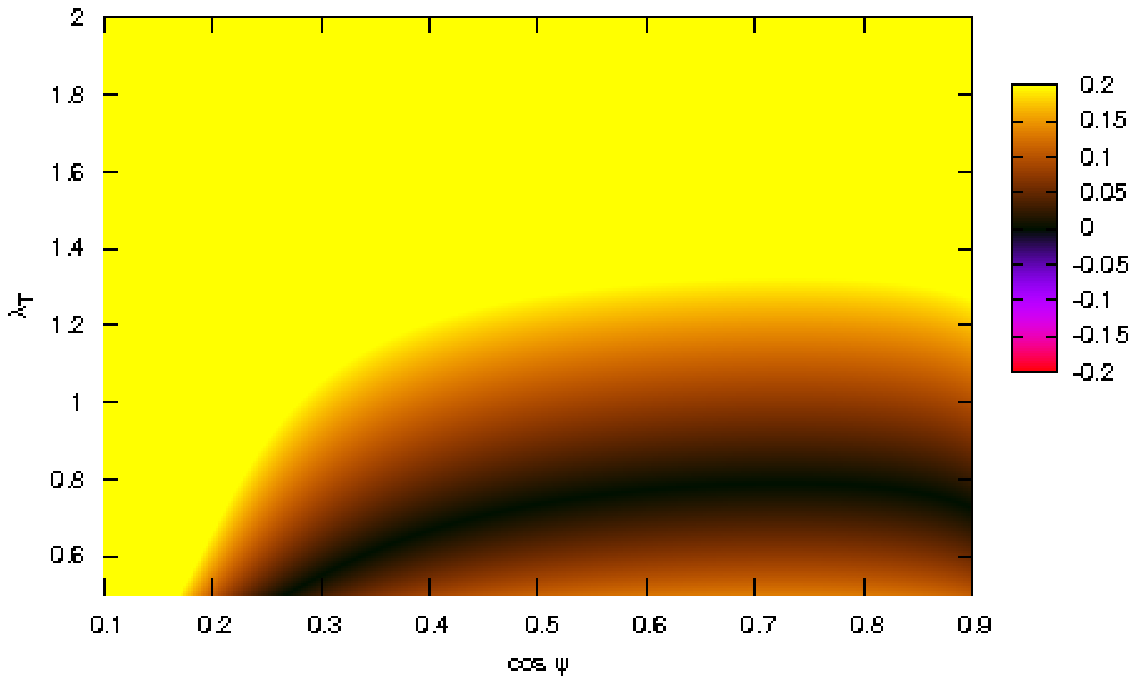,scale=0.4}\\
\end{tabular}
\end{center}\vspace*{-0.3cm}
\caption{Contours of the viable regions in the $\lambda_{T}-c{\psi}$ plane with the condition (\ref{cond})
 for different values of $a$ and $a'$. In the top and bottom rows $f$ and $\Lambda$ are fixed to $f=0.8$ TeV
 and $\Lambda=10$ TeV, and $f=1$ TeV and $\Lambda=12$ TeV, respectively.}
\label{CRyEO}
\end{figure}

In Fig.\ref{CRyEO} we show how the viable region changes with
different values of $f$, $\Lambda$, $a$ and $a'$. 
Deviations from the condition (\ref{cond}) are allowed up to $20 \%$.
In the top row
we have set the $f$ and $\Lambda$ values to $0.8$ TeV and $10$
TeV, and in the bottom row to $1$ TeV and $12$ TeV respectively.
Thus, the columns represent different  $a$ and $a'$ values.
Starting from left to right the values are the following: $a=0.3$
and $a'=0.2$, $a=0.3$ and $a'=0.7$, $a=0.9$ and $a'=0.2$, and
finally, $a=0.9$ and $a'=0.7$. One can see that small $a'$ values
are preferred in order to satisfy the SSB condition. Since in this
case we are taking into account both, the radiative corrections
and the effective operators,  the fermion sector contributions
become even more important than in the cases of considering some
of these contributions alone. Therefore,
the strongly dependence on $a'$  caused by the top's Yukawa
couplings, gives large values of the $\lambda$'s parameters for
high $a'$ values making difficult to satisfy the eq.(\ref{cond}).
However, the case of $a$ is quite different. From  Fig.\ref{CRyEO}
it is clear that the results do not depend strongly on this
parameter since its contribution is suppressed by the gauge
couplings $g$ and $g'$ and by the not so small values of
$c_{\psi}$.

It is interesting to note that if one consider only the contributions
to the potential coming from the effective operators (except for
the $\mu$ parameter which still will be dominated by the radiative
corrections), the SSB condition is not easily
satisfied. The reason is very simple; the $\lambda$'s parameters
at tree-level does not depend neither on $f$ nor on $\Lambda$,
while $\mu$ does. For example, the dependence of $\mu$ on $f$
appears trough the heavy particle masses and then $\mu$ is
directly proportional to $f$. In this way, $\mu$ rises very
quickly with $f$ meanwhile the $\lambda$'s parameters do not.
Therefore high values of the parameters $a$ and $a'$ are needed in
order to satisfy the condition (\ref{cond}). We find that
only for $f=0.8$ TeV, $\Lambda=10$ TeV, and $a\geq 0.75$ y $a'\geq
0.95$ the equality (\ref{cond}) is satisfied.  A more detailed
analysis of the allowed values of the constants $a$ and $a'$, in
agreement with the electroweak precision fits, is given
in~\cite{Csaki1}. 
The lowest value we found for $\mu$ is $\mu=0.49$ TeV, corresponding
to $f=0.8$ TeV, $\Lambda=10$ TeV, $c_{\psi}=0.2$ $\lambda=2.59$,
$a=0.75$ and $a'=0.95$, being the mass
of the $\phi$ scalar $2.86$ TeV.

From the discussion above we see that in all cases the  $\mu$
values are higher than  $350$ GeV. This is  far away from the
expected bound of the order of $200$ GeV predicted by the SM
precision tests. Therefore, it is clear that
the inclusion of the interactions terms
between GB and the other particles is not enough to reduce the
Higgs boson mass so that the complete compatibility with the
experimental constraints can be obtained. There are 
some indications suggesting that contributions coming from the scalar
sector must reduce the absolute value of $\mu^2$ and thus the
Higgs mass. Although the scalar loops contributions have not been
analyzed before (except the case of the radiatively generated 
scalar operators that have been discussed in~\cite{EspinosaNO}), 
the expression for the leading correction to the Higgs mass parameter, 
$\mu^{2}_{\phi}$, is presented in several articles. In particular, this correction 
is given by~\cite{Cohen},
\begin{equation}\label{muphi}
\mu^{2}_{\phi}= -\frac{\lambda}{16 \pi^{2}}\, M_\phi 
\log\left(1+\frac{\Lambda^{2}}{M_{\phi}^{2}}\right). 
\end{equation}
Let us now estimate the size of the above contribution for the case 
in which we have obtained the minimum
value for $\mu$,  $\mu=0.39$ TeV; corresponding to 
$f=0.8$ TeV, $\Lambda=10$ TeV, $\lambda_{T}=0.72$,
$c_{\psi}=0.34$ and $a=a'=0$, with $M_\phi=4.1$ TeV. 
Thus, by taking $\Lambda=10$ TeV, $M_\phi=4.1$ TeV and assuming
that $\lambda \simeq \frac{1}{3}$ (at tree level) for having a 
Higgs mass of order $200$ GeV~\cite{Csaki1}, we get $\mu_{\phi}=-0.14$ TeV. 
This implies that the $\mu$ value is reduced to be $\mu=250$ GeV. 
Note, however, that the quartic coupling $\lambda$ is fixed to a particular value 
in the above analysis. Since small changes in the input parameters 
of the model produce large changes in the value of $\lambda$ (and 
therefore the value of $\mu$ could vary), the radiative 
corrections to this coupling coming from the scalar sector must be also
taken into account in a full analysis~\cite{Future}. 

\section{Conclusions}

In this work we have computed the  Higgs and $\phi$ bosons
effective potential of the LH model  (\ref{potef1}).  We have
considered  two kind of contributions to the parameters of this
effective potential. Firstly, we have concentrated on the fermions
and gauge bosons one-loop radiative corrections. In this case, we
observe that the $\lambda$'s parameters are $O(\Lambda^2)$, since
they are not protected as the Higgs mass is in the LH model.
Secondly, we compute the contributions to the parameters  from the
effective operators coming from the ultraviolet completion of the
LH model. Here, the obtained parameters depend exclusively on the
two new unknown coefficients $a$ and $a'$, as well as the mixing
angle $c_{\psi}$ and the $T$ Yukawa coupling. This is an important
difference between the results obtained at one-loop which also
depend on the cutoff $\Lambda$.

The resulting potential has the right form to produce a
spontaneously symmetry breaking, since $\mu^2 > 0$ for some
regions of the LH model parameter space. Thus, if the
condition~(\ref{cond}) is satisfied, the electroweak symmetry is
broken. By using the obtained effective potential  we have
analyzed the constraints imposed on the LH parameters in order to
reproduce the SM electroweak symmetry breaking. We observe that if
one only consider the effective operator contribution, the SSB
condition is not easily satisfied. However, if the radiative
corrections or both contributions
  are taken into account the allowed ranges of the  parameters are much
  wider. The explanation comes from
  the way in which the coefficients of the potential, $\mu$ and $\lambda$'s,
   depend on $f$ and $\Lambda$, as it was discussed above.

Finally, we numerically analize the LH parameter space that can be set by 
requiring the LH Higgs effective potential to reproduce 
exactly the SM potential, and its compatibility with the present 
phenomenological constraints on the Higgs boson mass. 
The lowest value found for the $\mu$ parameter is
$390$ GeV, which implies a Higgs boson mass of $m_{H}\simeq 550$
GeV which is far from the current bound of about $200$ GeV. As a
consequence, we conclude that  radiative corrections, coming from
the Higgs itself and $\phi$ fields, could also provide relevant
contributions to the effective potential if the LH model is able
to reproduce the SM at low energies. 
An estimation of the scalar contribution to the $\mu$ parameter
leads to a value of $\mu=250$ GeV, and thus $m_H\simeq 350$ GeV. 
Nevertheless, the full contribution from the triplet, and thus the 
triplet mass $M_\phi$, is required to correct the Higgs mass in 
improved computations. The value of the Higgs quartic coupling,
$\lambda$, receives several contributions which have a non-trivial
dependence on the various parameters of the model and have no being 
computed so far. Work is in progress in order to compute these
contributions and to check if then the  value of Higgs mass will
get closer to the current bound~\cite{Future}.

\vspace{1cm}
{\bf Acknowledgments:} This work is supported by
DGICYT (Spain) under project number BPA2005-02327 and by the
Universidad Complutense/CAM project: number
 910309. The work of
S.P.\ has been supported by a {\it Ram{\'o}n y Cajal} contract
from MEC (Spain) and partially by CICYT (grant FPA2006-2315) and 
DGIID-DGA (grant 2008-E24/2). The work of L. Tabares-Cheluci is supported by
FPU grant from the Spanish M.E.C. We would like to thank
J.R.Espinosa and J.No Redondo for useful discussions.

\vspace{0.6cm}
{\bf{Appendix}}

The Goldstone bosons couplings to fermions and gauge bosons, needed for our computations,
are listened in the following:

\textbf{1. Couplings between Fermions and Goldstone Bosons}

\textit{a-Three particles}

$-\sqrt{2}\lambda_{T}H_{0}\overline{t}(1+\gamma_{5})T$

$-\sqrt{2}\lambda_{t}H_{0}\overline{t}(1+\gamma_{5})t$

$-\sqrt{2}\lambda_{T}H^{+}\overline{b}(1+\gamma_{5})T$

$-\sqrt{2}\lambda_{t}H^{+}\overline{b}(1+\gamma_{5})t$

\vspace{0.5cm}

\textit{b-Four particles}

\begin{tabular}{ll}
$-\frac{i\sqrt{2}}{f}\lambda_{T}H_{0}^{*}\phi_{0}\overline{t}(1+\gamma_{5})T$ \hspace{3cm}&\hspace{3cm}$-\frac{i\sqrt{2}}{f}\lambda_{t}H_{0}^{*}\phi_{0}\overline{t}(1+\gamma_{5})T$ \vspace*{0.3cm}\\
$-\frac{i}{f}\lambda_{T}H_{0}^{*}\phi^{+}\overline{b}(1+\gamma_{5})T$ \hspace{3cm}&\hspace{3cm}$-\frac{i}{f}\lambda_{t}H_{0}^{*}\phi^{+}\overline{b}(1+\gamma_{5})t$ \vspace*{0.3cm}\\
$-\frac{i}{f}\lambda_{T}H^{+*}\phi^{+}\overline{t}(1+\gamma_{5})T$ \hspace{3cm}&\hspace{3cm}$-\frac{i}{f}\lambda_{t}H^{+*}\phi^{+}\overline{t}(1+\gamma_{5})t$ \vspace*{0.3cm}\\
$-\frac{i}{f}\lambda_{T}H^{+*}\phi^{++}\overline{b}(1+\gamma_{5})T$ \hspace{3cm}&\hspace{3cm}$-\frac{i}{f}\lambda_{t}H^{+*}\phi^{++}\overline{b}(1+\gamma_{5})t$ \vspace*{0.3cm}\\
$-\frac{\lambda_{t}}{f}$tr$(\phi\phi^{\dag})\overline{T}(1+\gamma_{5})t$ \hspace{3cm}&\hspace{3cm}$-\frac{\lambda_{T}}{f}$tr$(\phi\phi^{\dag})\overline{T}T$
\end{tabular}

\vspace{0.5cm}

\textbf{2. Couplings between Gauge Bosons and Goldstone Bosons}

\textit{a-Four particles}

$\frac{g^{2}}{2}(\phi_{0}\phi_{0}^{*}+2\phi^{+}\phi^{-}+\phi^{++}\phi^{--})W_{1\mu}W_{1}^{\mu}$

$\frac{g^{2}}{2}(\phi_{0}\phi_{0}^{*}+2\phi^{+}\phi^{-}+\phi^{++}\phi^{--})W_{2\mu}W_{2}^{\mu}$

$2g^{2}(\phi_{0}\phi_{0}^{*}+\phi^{++}\phi^{--})W_{3\mu}W_{3}^{\mu}$

$\Big(-\frac{1}{2}\frac{g_{1}^{2}g_{2}^{2}}{g_{1}^{2}+g_{2}^{2}}$tr$(\phi\phi^{\dag})+\frac{1}{4}\frac{g_{1}^{4}+g_{2}^{4}}{g_{1}^{2}+g_{2}^{2}}\phi^{+}\phi^{-}\Big)W'_{1\mu}W_{1}^{'\mu}$

$\Big(-\frac{1}{2}\frac{g_{1}^{2}g_{2}^{2}}{g_{1}^{2}+g_{2}^{2}}$tr$(\phi\phi^{\dag})+\frac{1}{4}\frac{g_{1}^{4}+g_{2}^{4}}{g_{1}^{2}+g_{2}^{2}}\phi^{+}\phi^{-}\Big)W'_{2\mu}W_{2}^{'\mu}$

$\Big(-\frac{1}{2}\frac{g_{1}^{2}g_{2}^{2}}{g_{1}^{2}+g_{2}^{2}}$tr$(\phi\phi^{\dag})+\frac{1}{4}\frac{g_{1}^{4}+g_{2}^{4}}{g_{1}^{2}+g_{2}^{2}}(\phi_{0}\phi_{0}^{*}-\phi^{+}\phi^{-}+\phi^{++}\phi^{--})\Big)W'_{3\mu}W_{'3}^{\mu}$

$g^{'2}$tr$(\phi\phi^{\dag})b_{\mu}B^{\mu}$

$\frac{1}{4}\frac{(g_{1}^{'2}-g_{2}^{'2})^{2}}{g_{1}^{'2}+g_{2}^{'2}}$tr$(\phi\phi^{\dag})B'_{\mu}B^{'\mu}$

\vspace{0.5cm}

\textit{a-Five particles}

$\frac{i(g_{1}^{2}-g_{2}^{2})}{8f}(H_{0}\phi^{--}H_{0}+H^{+}\phi_{0}^{*}H^{+}+\sqrt{2}H_{0}\phi^{-}H^{+})W'_{1\mu}W_{1}^{'\mu}$+h.c.

$-\frac{i(g_{1}^{2}-g_{2}^{2})}{8f}(H_{0}\phi^{--}H_{0}+H^{+}\phi_{0}^{*}H^{+}-\sqrt{2}H_{0}\phi^{-}H^{+})W'_{2\mu}W_{2}^{'\mu}$+h.c.

$\frac{i(g_{1}^{2}-g_{2}^{2})}{8f}(H_{0}\phi^{--}H_{0}+H^{+}\phi_{0}^{*}H^{+}-\sqrt{2}H_{0}\phi^{-}H^{+})W'_{3\mu}W_{3}^{'\mu}$+h.c.

$\frac{i(g_{1}^{'2}-g_{'2}^{2})}{8f}(H_{0}\phi^{--}H_{0}+H^{+}\phi_{0}^{*}H^{+}+\sqrt{2}H_{0}\phi^{-}H^{+})B'_{\mu}B^{'\mu}$+h.c.

\end{document}